\documentclass[12pt]{report}
\usepackage[hscale=0.7, vscale=0.75]{geometry}
\usepackage{amsmath}
\usepackage{amssymb}
\usepackage{amsthm}
\usepackage{enumerate}
\usepackage{pdfpages}
\usepackage{graphicx}
\usepackage{array}

\usepackage{hyperref}  
\allowdisplaybreaks 

\newtheorem{dfn}{Definition}

\begin{document}
\title{\Large \textbf{An Overview of the Relationship between Group Theory and Representation Theory to the Special Functions in Mathematical Physics}\vspace{20pt}}
\author{Ryan D. Wasson \vspace{10pt} \\Advisor: Robert Gilmore, PhD \vspace{60pt} \\ A Thesis Submitted \\ in Partial  Fulfillment \\ of the Requirements \\ for the Degree of \\ Bachelor of Science\\ in Physics \vspace{25pt}\\ \textit{Drexel University,} \\ \textit{Philadelphia, PA}}

\date{\vspace{\fill}May 24, 2013}
\maketitle

\tableofcontents
\addtocontents{toc}{\protect\thispagestyle{empty}}
\thispagestyle{empty}

\chapter*{Acknowledgements}
\thispagestyle{empty}

This thesis would not have been possible without the help of my advisor, Dr.\ Robert Gilmore.  I would like to thank him for suggesting such a fascinating topic for my thesis and for his patience in mentoring me and helping me to grasp the ideas.  Thank you for encouraging me to keep reading, knowing that I would not understand anything in the beginning, and for taking such a keen interest in my thesis.  I really enjoyed our conversations and looked forward to them each week.  They were tremendously helpful for understanding the big picture.

I would also like to thank Dr.\ Robert Boyer for allowing me attend his graduate course on Lie groups this spring.  The lectures were helpful for understanding the mathematical background when writing Chapter \ref{ch:background}.

Finally, I would like to thank my family for their love and support throughout my time at Drexel. 

\begin{abstract}
Advances in mathematical physics during the 20th century led to the discovery of a relationship between group theory and representation theory with the theory of special functions.  Specifically, it was discovered that many of the special functions are (1) specific matrix elements of matrix representations of Lie groups, and (2) basis functions of operator representations of Lie algebras.  By viewing the special functions in this way, it is possible to derive many of their properties that were originally discovered using classical analysis, such as generating functions, differential relations, and recursion relations.  This relationship is of interest to physicists due to the fact that many of the common special functions, such as Hermite polynomials and Bessel functions, are related to remarkably simple Lie groups used in physics. 
Unfortunately, much of the literature on this subject remains inaccessible to undergraduate students.  The purpose of this project is to research the existing literature and to organize the results, presenting the information in a way that can be understood at the undergraduate level.  The primary objects of study will be the Heisenberg group and its relationship to the Hermite polynomials, as well as the Euclidean group in the plane and its relationship to the Bessel functions.  The ultimate goal is to make the results relevant for undergraduate students who have studied quantum mechanics.
\end{abstract}



\begin{figure}
\centering
\includegraphics{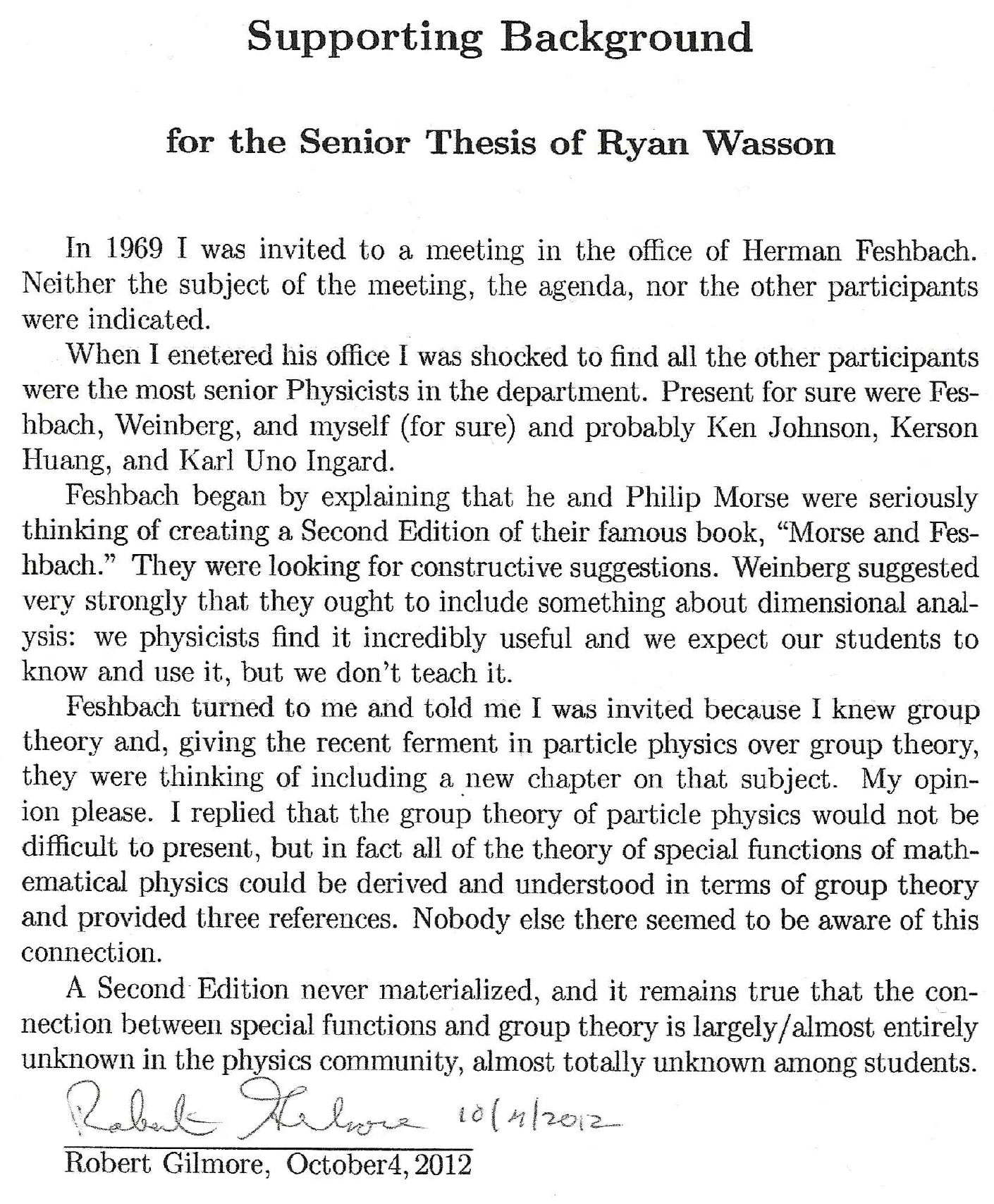}
\end{figure}
\thispagestyle{empty}

\clearpage
\setcounter{page}{1}

\chapter{Introduction}\label{ch:introduction}
Special functions are of primary importance to the study of physics.  They appear in physics as solutions to various differential equations describing many different kinds of physical systems.  In fact, one cannot study quantum mechanics without being exposed to many different special functions.  Hermite polynomials, e.g., appear in the solution to the quantum harmonic oscillator problem.  Bessel functions and Legendre polynomials appear when solving the Schr\"odinger equation in spherical coordinates.  The Laguerre polynomials appear when studying the hydrogen atom.  Despite the differences between the types of special functions that exist, they all share similar properties.  Special functions are orthogonal, satisfy various differential and recursion relations,  have generating functions, and are solutions to certain second order differential equations.

These properties were originally discovered during the 19th century by mathematicians doing classical analysis.  But during the middle of the 20th century, mathematicians found a relationship between the special functions and Lie groups.  This breakthrough allowed for a deeper understanding of the origin of the various properties of the special functions.  Lie groups are continuous groups with differentiable composition and inversion maps.  Physicists are most familiar with Lie groups having matrix representations, particularly in particle physics, e.g., $SU(2)$ or $SU(3)$.  It was discovered that many of the special functions are matrix elements of matrix representations of different Lie groups.  Many of the special functions also appear as basis functions of differential operator representations of Lie algebras.  A Lie algebra can be constructed from a Lie group by linearizing the group in the neighborhood of the identity.  The resulting Lie algebra is a linear vector space satisfying certain properties.  Rather than work directly with a Lie group, it is often easier to consider its corresponding Lie algebra instead.

The connection between Lie algebras and special functions had its beginnings in a 1951 paper by the mathematicians Infeld and Hull \cite{Infeld and Hull}.  Building on the efforts of physicists Paul Dirac and Erwin Schr\"odinger, who introduced a technique in quantum mechanics called the factorization method, Infeld and Hull discovered a way to factorize a large class of second order differential equations into pairs of first order differential operators.  This discovery provided an alternative way to derive recursion relations of the solutions to the differential equations. The equations they considered included many of the special function equations, such as Hermite's equation and Bessel's equation.  In the early 1960s, other mathematicians, including an American mathematician named Willard Miller \cite{Miller article}, were able to demonstrate that the first order differential operators that appear as factors in Infeld and Hull's paper are elements of Lie algebras.  The special functions and their properties could then be viewed as having their origin in the theory of Lie algebras and representation theory. 

\newcolumntype{A}{>{\centering\arraybackslash}m{3in}}
\newcolumntype{B}{>{\centering\arraybackslash}m{2in}}
\renewcommand{\arraystretch}{1.2}   
\begin{table}[t]
  \centering
  \begin{tabular}{| A | B |}
  \hline
 \textbf{Lie Group} & \textbf{Special Function(s)}  \\ \hline
 Heisenberg group $H_3$ & Hermite polynomials \\ \hline
 Euclidean group $E_2$ & Bessel functions \\ \hline
 Special unitary group $SU(2)$ & Legendre polynomials, \newline Jacobi polynomials \\ \hline
Unimodular quasi-unitary group $QU(2)$ & Legendre functions, \newline Jacobi functions \\ \hline
 Special linear group $SL(2,R)$ & Hypergeometric function \\ \hline
 Third order triangular matrices & Whittaker functions, \newline Laguerre polynomials \\ \hline
 Special orthogonal group $SO(n)$ & Gegenbauer functions \\
 \hline
  \end{tabular}
  \caption{The relationship between some of the special functions and various Lie groups \cite{Vilenkin book}.}
  \label{tab:relationship}
\end{table}
\renewcommand{\arraystretch}{1.0}   

In the years that followed, detailed treatments on the relationship between Lie groups and special functions have been given.  In 1968, three monographs on the subject were published.  Miller published a book called ``Lie Theory and Special Functions," which expanded on his earlier work  \cite{Miller book}.   James Talman published a book for physicists called ``Special Functions: A Group Theoretic Approach" which was based on lectures given by the physicist Eugene Wigner \cite{Talman book}.  Finally, the Russian mathematician N. Vilenkin published the book ``Special Functions and the Theory of Group Representations" (written in 1965; translated into English in 1968) \cite{Vilenkin book}.  While the focus in Miller's book is still partly on the factorization method and the relationship between special functions and representations of Lie algebras, all three of these books explore the relationship between special functions and representations of Lie groups. 

Despite the fact that undergraduate physics curriculum exposes students to some of the special functions and their properties, the connection between Lie theory and special functions is not taught.  Not only that, but many undergraduate physics students have no knowledge of what a Lie group even is.  This is unfortunate because, particularly in quantum mechanics, for example, there are Lie groups and Lie algebras present behind the scenes of many of the systems studied.  A knowledge of this connection would help students appreciate a deeper sense of some of the common problems in quantum mechanics, such as the quantum harmonic oscillator.  

On the other hand, the connection between special functions and Lie theory is an abstract subject requiring knowledge of advanced mathematics in order to grasp the basic ideas.  Much of the existing literature on the subject remains inaccessible to undergraduate students.  The purpose of this thesis is to bridge this divide and to make the subject relevant to undergraduate students who have studied quantum mechanics and are comfortable with bra-ket notation.  The necessary mathematical background is presented in Chapter \ref{ch:background}.  In Chapter \ref{ch:hermite polynomials}, the relationship between the Heisenberg algebra and the Hermite polynomials is explored.  The methods presented in Chapter \ref{ch:hermite polynomials} are then applied in Chapter \ref{ch:bessel functions} to study the relationship between the algebra of the Euclidean group in the plane and the Bessel functions.

\chapter{Background}\label{ch:background}
\section{Lie Groups}
Before one can begin to understand what a Lie group is, it is necessary to define the mathematical structure known as a \textit{group}.
\begin{dfn}
A \textbf{group} $(G,\circ)$ is a set $G$ together with a binary operation $\circ$ that satisfies the following four axioms:
\begin{enumerate}
\item $g_1 \circ g_2 \in G$ (Closure),
\item $(g_1 \circ g_2) \circ g_3 = g_1 \circ (g_2 \circ g_3)$ (Associativity),
\item There exists $e \in G$ such that $g_1\circ e = e\circ g_1 = g_1$ (Identity element),
\item There exists $g_1^{-1} \in G$ such that $g_1 \circ g_1^{-1} = g_1^{-1} \circ g_1= e$ (Inverse element)
\end{enumerate}
for any $g_1, g_2, g_3 \in G$. 
\end{dfn}
Note that associativity cannot be deduced from axioms $1, 3$ and $4$, and so it must be taken as an axiom.  There exist many examples of non-associative binary operations.  Subtraction is non-associative, for example, as is division and exponentiation. 

Groups are very common structures in mathematics.  Simple examples include the set of integers under addition $(\mathbb{Z},+)$, the set of real numbers excluding zero under multiplication $(\mathbb{R}\setminus\{0\},\times)$, and the set of complex $n \times n$ invertible matrices under matrix multiplication $(M_n(\mathbb{C}), \times)$.  If the binary operation of a group is clear, we often drop the operator symbol and write $g_1g_2$ in place of $g_1 \circ g_2$.    The number of elements in a group (called its order) can be finite or infinite.  

It is often convenient to parametrize the elements of the group using another set.  A parametrization of a group is essentially a method of labeling the elements of the group.  If we write the group elements in the form $g_i$ or $g(i)$ with $i \in \mathbb{N}$, then the indices $i$ serve as a label of the group.  Labeling the elements of a group with positive integers, however does not work for all groups.  If the order of a group is uncountably infinite, for example, then we must use another set (such as $\mathbb{R}^n$) to parametrize the group elements.

Once we have parametrized the elements of a group $G$ by a set $M$, we can view the group operation $\circ$ as a function $\phi$ from the direct product $M \times M$ to $M$, i.e., $\phi : M \times M \rightarrow M$.  Similarly, we can view the inversion property of the group $g^{-1}(x)=g(y)$ with $x,y\in M$ as a function $\psi: M \rightarrow M$.  As an example of this, if $M=\mathbb{R}$ is a parametrization of $G$, then group operations take the form $g(x) \circ g(y) = g(z)$ for $x, y, z \in \mathbb{R}$.  This is equivalent to writing $\phi(x,y) = z$, so here we have $\phi : \mathbb{R}\times \mathbb{R} \rightarrow \mathbb{R}$.  The inversion map takes the form $\psi: \mathbb{R} \rightarrow \mathbb{R}$ defined by $\psi(x)=y$.  

In the special case when $G$ is parametrized by an n-dimensional \textit{manifold}\footnote{A manifold is a topological space that resembles a Euclidean space ($\mathbb{R}^n$) on small scales.} 
 $M^n$ and the maps $\phi$ and $\psi$ are differentiable functions, then $G$ is a \textit{Lie group} \cite{Gilmore book}.
\begin{dfn}
A group $(G,\circ)$ parametrized by an n-dimensional manifold $M^n$ is a \textbf{Lie group} if its composition map $\phi:M^n\times M^n \rightarrow M^n$ and inversion map $\psi: M^n\rightarrow M^n$ are differentiable.
\end{dfn} 
An example of a Lie group relevant to this discussion is the \textit{Heisenberg group}, or $H_3$.  This group has a $3 \times 3$ matrix representation given by 
\begin{equation}\label{H_3 group matrix rep}
g(\textbf{x}) = \begin{pmatrix}
1 & x_1 & x_2 \\
0 & 1 & x_3 \\
0 & 0 & 1
\end{pmatrix} \in H_3
\end{equation}
with $\textbf{x}=(x_1, x_2, x_3) \in \mathbb{R}^3$ \cite{Gilmore article}.  It is easy to verify that $\phi$ and $\psi$ are differentiable maps, and so $H_3$ is indeed a Lie group.  For example, if $\textbf{y}=(y_1,y_2,y_3)$, then 
\begin{equation*}
g(\textbf{x}) \circ g(\textbf{y}) = 
\begin{pmatrix}
1 & x_1 & x_2 \\
0 & 1 & x_3 \\
0 & 0 & 1
\end{pmatrix}\begin{pmatrix}
1 & y_1 & y_2 \\
0 & 1 & y_3 \\
0 & 0 & 1
\end{pmatrix} = \begin{pmatrix}
1 & x_1+y_1 & y_2+x_1y_3+x_2 \\
0 & 1 & x_3+y_3 \\
0 & 0 & 1
\end{pmatrix}
\end{equation*}
and so $\phi(\textbf{x},\textbf{y}) = (x_1+y_1,y_2+x_1y_3+x_2,x_3+y_3)$ which is clearly differentiable in the group parameters.  It can similarly be shown that the inversion map $\psi(\textbf{x}) = (-x_1,x_1x_3-x_2,-x_3)$ is also differentiable.  

It is important to note that the composition and inversion maps are what define a Lie group, not a particular representation.  The Heisenberg group, for instance, is not defined by (\ref{H_3 group matrix rep}).  For a given Lie group, there are infinitely many ways to represent it.  We will discuss the subject of representation theory in section \ref{section Rep Theory}.


\section{Lie Algebras}

Whereas Lie groups are a special kind of group, Lie algebras are a special kind of linear vector space.  Lie algebras are closely related to Lie groups, in that when you linearize a Lie group in the neighborhood of its identity, you obtain a Lie algebra.  Conversely, when you exponentiate each element of a Lie algebra, you obtain elements of the Lie group.  The linearity of a Lie algebra makes it simpler to work with than its corresponding Lie group.  The commutation relations between basis vectors of the Lie algebra provide information about multiplication in the Lie group \cite{Gilmore book}. 

\begin{dfn}[cf.\ \cite{Miller article}]\label{dfn lie algebra}
A \textbf{Lie algebra} is an n-dimensional linear vector space $V$ over a scalar field $\mathbb{F}$ together with a binary operation $[\cdot,\cdot ]$ satisfying the following three axioms:
\begin{enumerate}
\item $[\alpha \textbf{u}+\beta \textbf{v}, \textbf{w} ] = \alpha [ \textbf{u}, \textbf{w}] + \beta [\textbf{v}, \textbf{w}]$ and $[\textbf{u},\alpha \textbf{v} + \beta \textbf{w}] = \alpha[\textbf{u},\textbf{v}]+\beta[\textbf{u},\textbf{w}]$ (Bilinearity),
\item $[\textbf{u},\textbf{v}] = -[\textbf{v},\textbf{u}]$ (Antisymmetry),
\item $[\textbf{u},[\textbf{v},\textbf{w}]]+[\textbf{v},[\textbf{w},\textbf{u}]]+[\textbf{w},[\textbf{u},\textbf{v}]]=0$ (Jacobi Identity)
\end{enumerate}
for any $\alpha, \beta \in \mathbb{F}$, $\textbf{u},\textbf{v}, \textbf{w} \in V$.
\end{dfn}

Both the cross product and the commutator operator satisfy these axioms.  Just as the concept of an inner product in a general vector space generalizes the concept of the dot product for vectors in Euclidean space, the vector product $[ \cdot, \cdot ]$ (called the Lie bracket) generalizes the concept of commutation.  A Lie algebra is essentially just a linear vector space with a vector product that behaves like the cross product or commutator operator.  Hence, simple examples of Lie algebras include the vector space $\mathbb{R}^3$ under the cross product and the vector space of $n\times n$ matrices $M_n(\mathbb{R})$ under the commutator operator.  

In general, a set of commutation relations between basis operators is what defines the Lie algebra \cite{Gilmore book}.  For example, the three-dimensional Heisenberg algebra $\mathfrak{h}_3$ with basis operators $X, Y, Z$ is defined by the commutation relations 
\begin{equation}\label {h_3 general com rel}
[X,Y]=[Y,Z]=0, \,\,\,\,[X,Z]=Y.
\end{equation}
The operators $X, Y, Z$ are elements of some vector space $V$.  To help perform computations and visualize the structure of the algebra, it is helpful to use a representation of $\mathfrak{h}_3$.  A $3 \times 3$ faithful matrix representation of $\mathfrak{h}_3$ is given by taking as basis operators 
 \begin{equation}\label{h_3 basis matrices}
A=\begin{pmatrix}
0 & 1 & 0 \\
0 & 0 & 0 \\
0 & 0 & 0 
\end{pmatrix} , \,\,\,\,B = 
\begin{pmatrix}
0 & 0 & 1 \\
0 & 0 & 0 \\
0 & 0 & 0 
\end{pmatrix} , \,\,\,\,C = 
\begin{pmatrix}
0 & 0 & 0 \\
0 & 0 & 1 \\
0 & 0 & 0 
\end{pmatrix} 
\end{equation}
since these matrices obey the commutation relations
\begin{equation}\label{h_3 commutation rel}
[A,B]=[B,C]=0, \,\,\,\,[A,C]=B.
\end{equation}
These commutations relations are identical to (\ref{h_3 general com rel}) if we make the association $X \rightarrow A$, $Y \rightarrow B$, $Z \rightarrow C$.
 An arbitrary element of the Heisenberg algebra can then be written 
\begin{equation}\label{h_3 matrix rep}
\begin{pmatrix}
0 & a & b \\
0 & 0 & c \\
0 & 0 & 0 
\end{pmatrix} = aA + bB + cC
\end{equation}
with $a, b, c \in \mathbb{R}$ \cite{Gilmore article}.

The Heisenberg algebra is closely related to the Heisenberg group.  Every Lie group is related to a particular Lie algebra since the set of all tangent vectors (i.e., derivatives) at the identity of curves in the group that pass through the identity form a Lie algebra.  The Heisenberg algebra is the set of all tangent vectors at the identity of curves in the Heisenberg group.  Using the $3\times 3$ matrix representation of $\mathfrak{h}_3$, we can easily illustrate this fact.  The basis vectors $A$, $B$, and $C$, for instance, are the tangent vectors
\begin{equation*}
\frac{\partial g(\textbf{x})}{\partial x_1}\Big|_{\textbf{x}=0} = \begin{pmatrix}
0 & 1 & 0 \\
0 & 0 & 0 \\
0 & 0 & 0 
\end{pmatrix}=A,
\end{equation*}
\begin{equation*}
\frac{\partial g(\textbf{x})}{\partial x_2}\Big|_{\textbf{x}=0} = \begin{pmatrix}
0 & 0 & 1 \\
0 & 0 & 0 \\
0 & 0 & 0 
\end{pmatrix}=B,
\end{equation*}
\begin{equation*}
\frac{\partial g(\textbf{x})}{\partial x_3}\Big|_{\textbf{x}=0} = \begin{pmatrix}
0 & 0 & 0 \\
0 & 0 & 1 \\
0 & 0 & 0 
\end{pmatrix}=C,
\end{equation*}
called the \textit{infinitesimal generators} of the representation.  Here $g(\textbf{x})$ is defined by (\ref{H_3 group matrix rep}).

Once you have constructed a Lie algebra from a Lie group, you can map operators in the algebra back to points in the group using the exponential map.  The exponential of a matrix $M$ is defined by the Taylor series expansion
\begin{equation*}
\exp(M) := \sum_{n=0}^\infty \frac{M^n}{n!}.
\end{equation*}
Using the $3\times 3$ matrix representation of the Heisenberg group and algebra, it is easy to show that matrices in $\mathfrak{h}_3$ get mapped to matrices in $H_3$, since in this case the  Taylor series expansion of the exponential is finite.   To see why, observe that
\begin{equation*}
 \begin{pmatrix}
0 & a & b \\
0 & 0 & c \\
0 & 0 & 0 
\end{pmatrix}^3=\begin{pmatrix}
0 & 0 & 0 \\
0 & 0 & 0 \\
0 & 0 & 0
\end{pmatrix},
 \end{equation*}
and so 
 \begin{align*}
 \exp \begin{pmatrix}
0 & a & b \\
0 & 0 & c \\
0 & 0 & 0 
\end{pmatrix} = I + \begin{pmatrix}
0 & a & b \\
0 & 0 & c \\
0 & 0 & 0 
\end{pmatrix} + \frac{1}{2} \begin{pmatrix}
0 & a & b \\
0 & 0 & c \\
0 & 0 & 0 
\end{pmatrix}^2 
= \begin{pmatrix}
1 & a & b+\frac{ac}{2} \\
0 & 1 & c \\
0 & 0 & 1
\end{pmatrix} \in H_3.
 \end{align*}

\section{Representation Theory}\label{section Rep Theory}

The $3 \times 3$ matrix representation of the Heisenberg algebra $\mathfrak{h}_3$ given in (\ref{h_3 matrix rep}) is not the only representation of $\mathfrak{h}_3$ possible.  In general, a given Lie algebra (and its Lie group) can have infinitely many possible representations.  The subject of \textit{representation theory} is devoted to studying and classifying different kinds of representations.  A \textit{representation} is essentially just a homomorphic\footnote{A homomorphism is a mapping $f:A\rightarrow B$ such that $f(xy) = f(x)f(y)$.  The exponential function is a homomorphism, for example.} mapping from elements of the Lie algebra (or Lie group) to a operators in a linear vector space.  First, we define a \textit{Lie group representation}.
\begin{dfn}\label{lie group rep}
A \textbf{representation} of a Lie group $G$ is a group $\Gamma$ of linear operators (e.g., matrices)  acting on a linear vector space $V$, together with a mapping $T:G\rightarrow \Gamma$ such that $T(ab) = T(a)T(b)$ for every $a,b \in G$. The vector space $V$ is called the \textbf{representation space} \textnormal{\cite{Talman book}}.
\end{dfn}

It is important to note that the product $T(a)T(b)$ in the definition above is understood to be function composition as $T(a)$ and $T(b)$ are operators.  If $\Gamma$ is a group of matrices then this is just matrix multiplication.  If $\Gamma$ is not a group of matrices, a matrix representation of $G$ can always be constructed relative to a given basis of $V$.  For example, suppose that $\{|e_i\rangle:i=1,\ldots, n\}$ is a set of basis vectors in an n-dimensional vector space $V$.  Then the matrix elements $T_{ij}$ are defined to be the coefficients in the expansion in the basis $\{|e_i\rangle \}$ of the operator $T$ acting on a basis vector $|e_j\rangle$ \cite{Talman book}.  That is,
\begin{equation}\label{dfn matrix elements}
T(a)\,|e_j\rangle = \sum_{i=1}^n T(a)_{ij}\,|e_i\rangle = \sum_{i=1}^n |e_i\rangle \, T(a)_{ij}
\end{equation}
If $V$ is an infinite-dimensional Hilbert space, then the sum becomes an infinite series or an integral.  If we define an inner product $\langle \cdot| \cdot \rangle$ on $V$ and choose an orthonormal basis set $\{|e_i\rangle\}$, then the matrix elements $T_{ij}$ can be obtained by taking the inner product of both sides of (\ref{dfn matrix elements}) with respect to a basis vector $|e_i\rangle$:
\begin{equation} \label{matrix elements solved}
T(a)_{ij}=\langle e_i | T(a) | e_j \rangle
\end{equation}
This procedure is none other than what David Griffiths refers to as ``Fourier's trick" in Introduction to Quantum Mechanics \cite{Griffiths book}.

Next, we define a \textit{Lie algebra representation}.
\begin{dfn}\label{lie alg rep}
A \textbf{representation} of a Lie algebra $\mathfrak{g}$ is a Lie algebra $\gamma$ of linear operators (e.g., matrices) acting on a linear vector space $V$, together with a mapping $t: \mathfrak{g} \rightarrow \gamma$ such that $t([a,b])=[t(a),t(b)]= t(a)t(b)-t(b)t(a)$ for every $a,b \in \mathfrak{g}$.  The vector space $V$ is called the \textbf{representation space} \textnormal{\cite{Miller book}}.
\end{dfn}

In this definition, the product $t(a)t(b)$ is again taken to be function composition.  Also note that the vector product in $\gamma$ is assumed to be the commutator.  In general, the vector product of an arbitrary Lie algebra can always be defined differently (as long as it satisfies the three axioms in Definition \ref{dfn lie algebra}), but in order for a Lie algebra to be considered a representation of another Lie algebra, the former must use the commutator vector product.  

It is possible to construct a matrix representation of a Lie algebra in a manner similar to that described above for Lie group representations.  An example of a $3 \times 3$ faithful matrix representation of $\mathfrak{h}_3$ was already given above in (\ref{h_3 matrix rep}).   Note that a \textit{faithful} representation is one that is one-to-one and onto (i.e., bijective) \cite{Talman book}. 

It was stated in the introduction that special functions appear as matrix elements of matrix representations of different Lie groups and also as basis functions of differential operator representations of Lie algebras.  Not all representations of Lie groups and Lie algebras, however, are related to special functions.  For example, the $3 \times 3$ matrix representations of the Heisenberg group $H_3$ and its algebra $\mathfrak{h}_3$ given earlier are not directly related to special functions.  The special functions of interest in this thesis appear only when considering certain kinds of \textit{unitary} representations.  
\begin{dfn}
A representation $\Gamma$ of linear operators acting on a linear vector space $V$ is called \textbf{unitary} if for all $f,g \in V$ and $T \in \Gamma$ we have $\langle f | g \rangle = \langle Tf | Tg \rangle$.  If a matrix representation $\tilde{T}$ defined by \textnormal{(\ref{matrix elements solved})} is expressed using an orthonormal basis, then $\tilde{T}^\dagger \tilde{T} = I$ \textnormal{\cite{Vilenkin book}}.
\end{dfn}
This definition requires that an inner product $\langle \cdot | \cdot \rangle$ be defined on the vector space $V$.  When $\tilde{T}$ is expressed using an orthonormal basis, the matrix $\tilde{T}^\dagger$ is taken to be the Hermitian adjoint, or conjugate transpose, of the matrix $\tilde{T}$.

There are no nontrivial finite-dimensional unitary representations of $H_3$ or anti-unitary representations of $\mathfrak{h}_3$\footnote{Technically, this is because $H_3$ and $\mathfrak{h}_3$ are non-compact.}.  It is not until infinite-dimensional unitary representations of $H_3$ or $\mathfrak{h}_3$ are considered that special functions (the Hermite polynomials) appear.  

The fact that infinitely many possible representations exist for a given Lie group or algebra might make the subject of representation theory sound quite complicated.  Fortunately, there are ways to classify the different representations that simplify the theory substantially.  Different representations are deemed \textit{equivalent} if there exists a similarity transformation between them.  That is, two representations $T_1$ and $T_2$ are equivalent if there exists an operator $S$ such that $T_1 = ST_2S^{-1}$.  Representations are called \textit{irreducible} if they cannot be decomposed into simpler representations.  In a sense, irreducible representations are the building blocks of all representations much like prime numbers are the building blocks of all positive integers.  All of the representations of a given Lie group or Lie algebra can be constructed from irreducible representations \cite{Talman book}

Unitary representations are nice to work with because if they are not irreducible to begin with, they can always be written as a \textit{direct sum} of irreducible representations (this is referred to as being \textit{completely reducible}).  Direct sums in matrix form appear block diagonal.  Thus, a unitary representation written in matrix form is a block diagonal matrix consisting of irreducible representations on the diagonal \cite{Talman book}.


An example of a unitary irreducible differential operator representation of $\mathfrak{h}_3$ consists of the familiar ladder operators used to solve the quantum harmonic oscillator problem.  Recall that the ladder operators are defined by\footnote{This definition tacitly assumes a choice of basis.  In this case, the ladder operators have been expressed using the position basis $|x \rangle$.}
\begin{equation}\label{ladder operators}
a_- = \frac{1}{\sqrt{2}} \left( x + \frac{d}{dx} \right), \,\,\,\,a_+ = \frac{1}{\sqrt{2}} \left( x - \frac{d}{dx} \right).
\end{equation}
If we take $\gamma$ to be the Lie algebra with basis vectors $a_-, a_+$, and the identity operator $I$, we can define the mapping $t: \mathfrak{h}_3 \rightarrow \gamma$ by $t(X) = a_-$, $t(Y) = I$, and $t(Z) = a_+$.  It is easy to verify that $t$ is a representation of $\mathfrak{h}_3$ since the basis operators $a_-$, $a_+$, and $I$ obey the same commutation relations given in (\ref{h_3 general com rel}) for the operators $X$, $Y$, and $Z$.  We will show in the next section that this representation is related to the special functions known as the Hermite polynomials.

\chapter{Hermite Polynomials}\label{ch:hermite polynomials}

The properties of the Hermite polynomials can be derived using the 
differential operator representation of the Heisenberg algebra $\mathfrak{h}_3$ given in (\ref{ladder operators}).  These operators are the familiar ladder operators used to solve the quantum harmonic oscillator problem.  In this chapter, we will show that the problem of deriving the Hermite polynomials essentially reduces to the quantum harmonic oscillator problem.  We will then use these operators to derive the properties listed in Appendex \ref{sec:hermite table}.  

\section{The Hermite Polynomials as Basis Functions of the Representation Space}\label{sec:hermite as basis functions}

The ladder operators $a_\pm$ act on a linear vector space of continuous ($C^\infty$) functions of one variable, called the \textit{representation space}.  This vector space has two standard bases:  the continuous (position) basis, with basis vectors denoted $|x\rangle$, and the discrete (energy) basis, with basis vectors denoted $|n\rangle$.  The continuous basis provides us with a geometric description of the representation space, while the discrete basis provides us with an algebraic description.  

In this section\footnote{The methods used in sections \ref{sec:hermite as basis functions} - \ref{sec:hermite gen func}  are outlined in \cite{Gilmore article}.} we show that  the Hermite polynomials are proportional to the mixed basis functions $\langle x | n \rangle$.  The mixed basis functions are nothing more than the coefficients of the projection of the discrete basis into the continuous basis.  Hence, the Hermite polynomials have their origin in both the geometric description of the representation space and the algebraic description.  We could also view them as matrix elements of the identity operator in the mixed basis since $\langle x | n \rangle = \langle x | I | n \rangle$.  

Before getting started, it is useful to write down a few of the properties of the continuous and discrete basis functions.  The continuous basis functions $|x\rangle$ are orthonormal and complete, and so
\begin{equation}\label{continuous}
\langle x' | x \rangle = \delta(x'-x),\;\;\;\;\int_{-\infty}^{\infty} |x\rangle \langle x| \, dx = 1.
\end{equation}
Hence, in this basis, the matrix elements of the basis operators $a_-$, $a_+$ and $I$ are
\begin{equation}\label{eq:continuous matrix elements}
\begin{aligned}
\langle x' | a_- | x \rangle &= \frac{1}{\sqrt{2}} \left( x + \frac{d}{dx} \right) \delta( x' - x ),\\
\langle x' | a_+ | x \rangle &= \frac{1}{\sqrt{2}} \left( x - \frac{d}{dx} \right) \delta( x' - x ),\\
\langle x' | I | x \rangle &= \delta( x' - x ) .
\end{aligned}
\end{equation}

The discrete basis functions $|n\rangle$ are also orthonormal and complete, hence
\begin{equation}\label{discrete}
\langle n' | n \rangle = \delta_{n',n}, \,\,\,\, \sum_{n=0}^\infty |n\rangle \langle n| = 1.
\end{equation}
From quantum mechanics \cite{Griffiths book} we know that these operators (\ref{ladder operators}) satisfy
\begin{equation}\label{discrete operators}
a_- |n\rangle = \sqrt{n} |n-1\rangle, \,\,\,\, a_+ |n\rangle = \sqrt{n+1} |n+1\rangle.
\end{equation}
Thus, an infinite-dimensional matrix representation of these operators in the discrete basis is given by
\begin{equation}\label{eq:discrete matrix elements}
\begin{aligned}
\langle n' | a_- | n \rangle & = \langle n' | n-1 \rangle \sqrt{n} = \delta_{n', n-1} \sqrt{n}, \\
\langle n' | a_+ | n \rangle& = \langle n' | n+1 \rangle \sqrt{n+1} = \delta_{n', n+1} \sqrt{n+1},\\
\langle n' | I | n \rangle &= \langle n' | n \rangle = \delta_{n',n}.
\end{aligned}
\end{equation}

In order to compute the mixed basis functions $\langle x | n \rangle$, we first compute the zeroth basis function $\langle x | 0 \rangle$.  We then proceed to apply the raising operator $a_+$ to this first state to obtain the general mixed basis function $\langle x | n \rangle$, just as one would do when solving the quantum harmonic oscillator problem.  The zeroth basis function is obtained by computing the matrix elements 
\begin{equation}\label{eq: a_- mixed matrix 0}
\langle x | a_- | 0 \rangle 
\end{equation}
of the $a_-$ operator in the mixed basis.  Since the matrix elements of $a_-$ with respect to the continuous basis are well defined, we can compute (\ref{eq: a_- mixed matrix 0}) by inserting the completeness property of the continuous basis into the bra-kets:
\begin{align*}
\langle x | a_- | 0 \rangle = \int_{-\infty}^{\infty} \langle x | a_- | x' \rangle \langle x' | 0 \rangle \, dx' &= \int_{-\infty}^{\infty} \left[ \frac{1}{\sqrt{2}} \left( x + \frac{d}{dx} \right) \delta( x' - x ) \right] \langle x' | 0 \rangle \, dx' \\
& = \frac{1}{\sqrt{2}} \left( x + \frac{d}{dx} \right) \langle x | 0 \rangle.
\end{align*}
This procedure is known as a \textit{resolution of the identity}.  Alternatively, we can perform a resolution of the identity for the discrete basis and use the fact that the matrix elements of $a_-$ with respect to the discrete basis are well defined:
\begin{equation*}
\langle x | a_- | 0 \rangle = \sum_{n=0}^{\infty} \langle x | n \rangle \langle n | a_- | 0 \rangle = 0.
\end{equation*}
  
Equating these two expressions results in a simple ODE we can solve:
\begin{equation*}
\frac{1}{\sqrt{2}} \left( x + \frac{d}{dx} \right) \langle x | 0 \rangle = 0 \;\;\;\; \Rightarrow \;\;\;\; \frac{d\psi_0}{dx} = -x \psi_0.
\end{equation*}
Solving for $\psi_0 = \langle x | 0 \rangle$, we find that $\langle x | 0 \rangle = Ce^{-x^2/2}$ for some constant C.  By normalizing this function so that its square integrates to one over the real line (giving $C = \pi^{-1/4}$), we obtain the ground state wave function of the harmonic oscillator. 

We can now compute the general matrix elements $\langle x | n \rangle$ by apply the raising operator $a_+$ to $|0 \rangle$ $n$ times:
\begin{equation*}
\langle x | n \rangle = \langle x | \frac{(a_+)^n}{\sqrt{n!}} | 0 \rangle.
\end{equation*}
Using a resolution of the identity and (\ref{hermite rod}),
\begin{align}
\langle x | n \rangle &= \int_{-\infty}^{\infty} \langle x | \frac{(a_+)^n}{\sqrt{n!}} | x' \rangle \langle x' | 0 \rangle \, dx' \notag \\
&= \int_{-\infty}^{\infty} \left[ \frac{1}{\sqrt{n!2^n}} \left( x - \frac{d}{dx} \right)^n \delta( x' - x )\right] \left[ \frac{e^{-(x')^2/2}}{ \pi^{-1/4}}\right] \, dx' \notag \\
&= \frac{1}{\sqrt{n!2^n\sqrt{\pi}}} \left( x - \frac{d}{dx} \right)^n e^{-x^2/2} \notag \\
&=  \frac{e^{-x^2/2} H_n(x)}{\sqrt{n!2^n\sqrt{\pi}}}. \label{<x|n>}
\end{align}
Thus, the matrix elements $\langle x | n \rangle$ are proportional to the Hermite polynomials.  These are, in fact, the solutions to the quantum harmonic oscillator problem.  If this surprises you, recall that the wave function $\psi_n(x)$ in quantum mechanics can be written (cf.\ \cite{Griffiths book})
\begin{equation*}
\psi_n(x) = \langle x | \psi_n \rangle
\end{equation*}
using bra-ket notation.  The discrete basis functions $|n\rangle$ are equivalent to the energy eigenfunctions $\psi_n$.  

\begin{figure}[t!]
\centering
\includegraphics[width=0.795\textwidth]{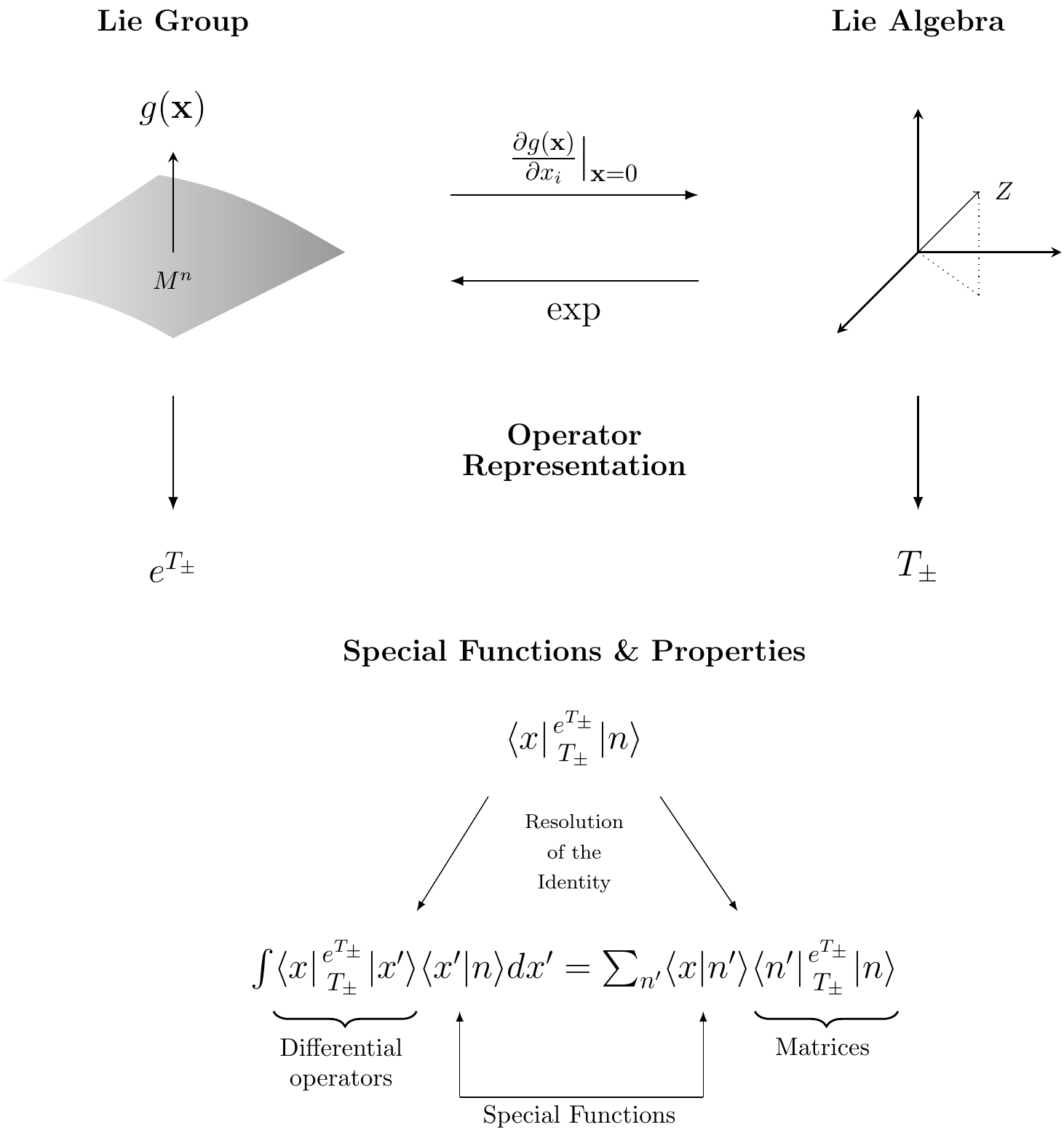}
\caption{The big picture.}
\label{fig:big picture}
\end{figure}

This example has illustrated a standard procedure for computing mixed matrix elements of an arbitrary operator $X$ in the Lie algebra.  We calculated the matrix elements of the operator $X = a_-$ with respect to each basis by performing a resolution of the identity and equating the results:
\begin{equation} \begin{array}{rcl}
 & \langle  x | X |  n \rangle & \\
\swarrow& & \searrow \\
\int \langle x | X | x'\rangle \langle x'| n \rangle\,dx'
 &=&
\sum_{n'} \langle { x} | { n'} \rangle \langle { n'} | X |{n}\rangle
\end{array}
\label{eq:standard procedure}
\end{equation}
This procedure will be used repeatedly throughout this thesis to derive properties of the Hermite polynomials, and in a later section, the Bessel functions.  The general relationship between Lie groups, Lie algebras, and special functions is summarized in Figure \ref{fig:big picture}.

\section{Differential Equation}

The differential equation (\ref{hermite diff eq}) satisfied by the Hermite polynomials can be obtained by computing the matrix elements of the anticommutator $\{a_-,a_+\}$ in the mixed basis, where $\{a_-,a_+\} = (a_-a_+ + a_+a_-)$.  Using the continuous basis, we have
\begin{align*}
\langle x | \{ a_-,a_+\} | n \rangle &= \int_{-\infty}^{\infty} \langle x | \{ a_-,a_+\} | x' \rangle \langle x' | n \rangle \, dx' \\
&=\{ a_-,a_+\} \int_{-\infty}^{\infty} \delta(x-x')\langle x' | n \rangle\,dx' \\
&=\{ a_-,a_+\} \langle x | n \rangle.
\end{align*} 
For convenience, let $D = \frac{d}{dx}$. Then by (\ref{ladder operators}),
\begin{equation*}
a_-a_+ = \frac{1}{2}\left(x+D\right)\left( x-D\right) = \frac{1}{2} (x^2 -D^2 +Dx -xD),
\end{equation*}
\begin{equation*}
a_+a_- = \frac{1}{2}\left(x-D\right)\left( x+D\right) = \frac{1}{2} (x^2 -D^2 -Dx +xD),
\end{equation*}
and so
\begin{equation*}
\{a_-,a_+\} = x^2-D^2. 
\end{equation*}
By (\ref{<x|n>}), the second derivative of $\langle x | n \rangle$ is proportional to
\begin{equation*}
D^2 (e^{-x^2/2} H_n(x))=e^{-x^2/2}(H_n''(x) - 2xH_n'(x)+(x^2-1)H_n(x)),
\end{equation*} 
and so
\begin{align}
\{a_-,a_+\} \langle x | n \rangle &= \beta( x^2 e^{-x^2/2} H_n(x) -D^2 (e^{-x^2/2} H_n(x))) \notag \\
&=\beta e^{-x^2/2} ( -H_n''(x) + 2xH_n'(x) + H_n(x) )   \label{hermite diff continuous}
\end{align}
with $\beta = (n!\,2^n \sqrt{\pi})^{-1/2}$.  But using the discrete basis, by (\ref{discrete operators}) we have
\begin{align*}
a_-a_+ \langle x | n \rangle &= a_- (\sqrt{n+1} \,\langle x | n+1\rangle) = (n+1) \langle x | n \rangle, \\
a_+a_-\langle x | n \rangle &= a_+(\sqrt{n}\,\langle x | n-1\rangle) = n \langle x | n \rangle,
\end{align*}
and so
\begin{align}
\{a_-,a_+\}\langle x | n \rangle &= (2n+1) \langle x | n \rangle \notag \\
&=(2n+1)\beta e^{-x^2/2} H_n(x).\label{hermite diff discrete}
\end{align}
Equating (\ref{hermite diff continuous}) and (\ref{hermite diff discrete}) and dividing by $\beta e^{-x^2/2}$ gives
\begin{equation*}
-H_n''(x) +2xH_n'(x)+H_n(x) = (2n+1)H_n(x).
\end{equation*}
After rearranging terms, we obtain the differential equation
\begin{equation*}
H_n''(x)-2xH_n'(x)+2nH_n(x)=0.
\end{equation*}

\section{Recursion Relation}\label{section hermite recursion}

We can derive the recursion relation (\ref{hermite recursion}) by computing the matrix elements of the operator $\hat{x} = (a_- + a_+) / \sqrt{2} \in \mathfrak{h}_3$ in the mixed basis:
\begin{align}
\frac{1}{\sqrt{2}} \langle x | a_-+a_+ | n \rangle & =  \frac{1}{\sqrt{2}} \left( \langle x | a_- | n \rangle + \langle x | a_+ | n \rangle \right)\notag \\
&= \frac{1}{\sqrt{2}} \left( \sqrt{n} \langle x | n-1 \rangle + \sqrt{n+1} \langle x | n+1 \rangle \right) .\label{recursion RHS}
\end{align}
But we can also write
\begin{align}\label{recursion LHS}
\langle x | \hat{x} | n \rangle &= \int_{-\infty}^{\infty} \langle x | \hat{x} | y \rangle \langle y | n \rangle \, dy \notag \\
& = x \int_{-\infty}^{\infty} \delta(x-y) \langle y | n \rangle \, dy \notag \\
&= x \langle x | n \rangle\notag \\
& = x \frac{e^{-x^2/2} H_n(x)}{\sqrt{n!2^n\sqrt{\pi}}}.
\end{align}
Equating (\ref{recursion RHS}) and (\ref{recursion LHS}) gives
\begin{align*}
  \frac{x H_n(x)}{\sqrt{n!\,2^n}} & =  \frac{1}{\sqrt{2}} \left( \frac{\sqrt{n} H_{n-1}(x)}{\sqrt{(n-1)!\,2^{n-1}} } + \frac{ \sqrt{n+1} H_{n+1}(x)}{\sqrt{(n+1)!\,2^{n+1}}} \right) \\
  \Rightarrow \;\;\;\;  \sqrt{2}\,xH_n(x) & =  n\sqrt{2}\,H_{n-1}(x)+\frac{H_{n+1}(x)}{\sqrt{2}} .
\end{align*}
Multiplying by $\sqrt{2}$ and rearranging terms gives (\ref{hermite recursion}).

\section{Differential Relation} \label{section hermite differential}
Similarly, we can derive the differential relation (\ref{hermite diff rel}) by computing the matrix elements of the operator $\frac{d}{dx} = (a_--a_+) / \sqrt{2}\in \mathfrak{h}_3$ in the mixed basis.  Compared with (\ref{recursion RHS}), the only difference is a minus sign:
\begin{equation*}
\frac{1}{\sqrt{2}} \langle x | a_--a_+ | n \rangle = \frac{1}{\sqrt{2}} \left( \sqrt{n} \langle x | n-1 \rangle - \sqrt{n+1} \langle x | n+1 \rangle \right).
\end{equation*}
Similar to (\ref{recursion LHS}), we have  
\begin{align*}
\langle x | \frac{d}{dx} | n \rangle = \frac{d}{dx} \langle x | n \rangle & = \frac{d}{dx} \left( \frac{e^{-x^2/2} H_n(x)}{\sqrt{n!\,2^n\sqrt{\pi}}} \right) \\
& = \frac{e^{-x^2/2}}{\sqrt{n!\,2^n\sqrt{\pi}}} \left( H_n'(x) -x H_n(x)\right).
\end{align*}
Thus,
\begin{align*}
\frac{H_n'(x) -x H_n(x)}{\sqrt{n!\,2^n}} &= \frac{1}{\sqrt{2}} \left( \frac{\sqrt{n} H_{n-1}(x)}{\sqrt{(n-1)!\,2^{n-1}} } - \frac{ \sqrt{n+1} H_{n+1}(x)}{\sqrt{(n+1)!\,2^{n+1}}} \right) \\
\Rightarrow \;\;\;\; H_n'(x) -x H_n(x) &= nH_{n-1}(x)-\frac{H_{n+1}(x)}{2} \\
\Rightarrow \;\;\;\; H_n'(x) &= 2n H_{n-1}(x)
\end{align*}
which is the differential relation in (\ref{hermite diff rel}).  We used the recursion relation in the last step.

\section{Generating Function}\label{sec:hermite gen func}
To derive the generating function (\ref{hermite gen func}), we compute the matrix elements of the exponential $e^{\sqrt{2} t a_+}$ in the mixed basis, with $n=0$.  This object lives in the Lie group ($H_3$) since it is the exponential of an operator in the Lie algebra ($\mathfrak{h}_3$).  First observe that we can write
\begin{align}\label{eq:hermite gen func discrete res}
\langle x | e^{\sqrt{2}ta_+} | 0 \rangle &= \sum_{n=0}^\infty \langle x | n \rangle \langle n | e^{\sqrt{2}ta_+} | 0 \rangle.
\end{align}
To compute the matrix elements $\langle n | e^{\sqrt{2}ta_+} | 0 \rangle$, we perform a Taylor series expansion of the exponential operator:
\begin{align*}
\langle n | e^{\sqrt{2}ta_+} | 0 \rangle &= \sum_{m = 0}^\infty \frac{(\sqrt{2}\, t)^m}{m!} \langle n |(a_+)^m|0\rangle \\
&= \sum_{m = 0}^\infty \frac{(\sqrt{2}\,t)^m}{m!}\sqrt{m!}\, \delta_{n,m}.
\end{align*}
The summand is nonzero only when $m=n$.  Hence, (\ref{eq:hermite gen func discrete res}) becomes
\begin{align}
\langle x | e^{\sqrt{2}ta_+} | 0 \rangle &= \sum_{n=0}^\infty \langle x | n \rangle \left( \frac{\sqrt{2^n n!}}{n!} t^n \right) \notag \\
&= \frac{e^{-x^2/2}}{\pi^{1/4}} \sum_{n=0}^{\infty} \frac{t^nH_n(x)}{n!} \label{gen func taylor}
\end{align}
by (\ref{<x|n>}).
Alternatively we can write (\ref{eq:hermite gen func discrete res}) as
\begin{equation*}
\langle x | e^{\sqrt{2}ta_+} | 0 \rangle = \int_{-\infty}^\infty \langle x | e^{\sqrt{2}ta_+} | x' \rangle \langle x' | 0 \rangle\,dx'.
\end{equation*}
Since (see \cite{Gilmore gen func}) 
\begin{equation*}
\langle x | e^{\sqrt{2}ta_+} | x' \rangle=\delta(x-e^{\sqrt{2}ta_+}x'),
\end{equation*}
we get that
\begin{align}
\langle x | e^{\sqrt{2}ta_+} | 0 \rangle &= e^{t(x-D)} \langle x | 0 \rangle \notag \\
&= e^{t(x-D)} \left(\frac{e^{-x^2/2}}{\pi^{1/4}}\right) \label{eq:before disentangling}
\end{align}
since $\sqrt{2} t a_+ = t(x-D)$, where $D = \frac{d}{dx}$.  

In order to evaluate this last expression, we must rewrite $e^{t(x-D)}$ using what are called ``disentangling" theorems.  Disentangling theorems provide the rules for factorizing exponentials of operators in Lie algebras.  Ordinarily, an exponential of the form $e^{a+b}$ equals $e^ae^b$ after factorization, with real numbers $a$ and $b$.   If $a$ and $b$ are non-commutative operators, however, then this is not necessarily correct.  In our case, $x$ and $D$ do not commute.  The disentangling theorem for $e^{t(x-D)}$ is
\begin{equation*}
e^{t(x-D)} = e^{tx}e^{-t^2/2}e^{-tD}.
\end{equation*}
We can use it to rewrite (\ref{eq:before disentangling}):
\begin{equation*}
\langle x | e^{\sqrt{2} t a_+} | 0 \rangle = e^{tx}e^{-t^2/2}e^{-tD}\left( \frac{e^{-x^2/2}}{\pi^{1/4}}\right).
\end{equation*}
But note that for any differentiable function $f(x)$,
\begin{align*}
e^{-tD}f(x) &= \sum_{n=0}^{\infty}\frac{(-t D)^n}{n!} f(x) \\
&= \sum_{n=0}^{\infty} \frac{f^{(n)}(x)}{n!}(-t)^n \\
&= f(x-t),
\end{align*}
and so
\begin{equation*}
\langle x | e^{\sqrt{2} t a_+} | 0 \rangle =\frac{1}{\pi^{1/4}} e^{tx}e^{-t^2/2}e^{-(x-t)^2/2}.
\end{equation*}
Equating this with (\ref{gen func taylor}) gives
\begin{equation*}
e^{2xt-t^2}= \sum_{n=0}^\infty \frac{H_n(x)}{n!} t^n,
\end{equation*}
which matches (\ref{hermite gen func}).

\chapter{Bessel Functions}\label{ch:bessel functions}

The program outlined in the preceding chapter is not limited to the Hermite polynomials.  Many other special functions are related to representations of simple Lie groups and their corresponding algebras.  In this chapter, we demonstrate the relationship between the Bessel functions and the Euclidean group in the plane, and we apply the same methods used in Chapter \ref{ch:hermite polynomials} to derive the properties listed in Appendix \ref{sec:bessel table}.  We conclude the chapter with a presentation of an alternative procedure for obtaining the Bessel functions, called \textit{group contraction}.  Note that the derivations in this chapter are presented in a more concise manner than was the case for the Hermite polynomials.  For more detailed derivations, consult Appendix \ref{app:bessel extra details}.

\section{The Euclidean Group in the Plane}

The Euclidean group in the plane, denoted $E_2$, consists of all possible translations and rotations in the Euclidean plane $\mathbb{R}^2$. It has a faithful $3 \times 3$ matrix representation given by
\begin{equation*}
g(x,y,\theta) = \begin{pmatrix}
\cos{\theta} & -\sin{\theta} & x \\
\sin{\theta} & \cos{\theta} & y \\
0 & 0 & 1
\end{pmatrix} \in E_2
\end{equation*}
with group parameters $x, y, \theta \in \mathbb{R}$ and $0 \leq \theta < 2\pi$.  In order to have group elements in $E_2$ operate on vectors in $\mathbb{R}^2$ using this representation, it is necessary to write an arbitrary vector $(a,b) \in \mathbb{R}^2$ as a vector $(a,b,1) \in \mathbb{R}^3$ \cite{Talman book}.  For example, let $\textbf{v} = (a,b) \in \mathbb{R}^2$.  To rotate $\textbf{v}$ by an angle $\theta$ and then to translate the resulting vector by the vector $(x,y)$, we can write:
\begin{equation*}
g(x,y,\theta) \, \textbf{v} = \begin{pmatrix}
\cos{\theta} & -\sin{\theta} & x \\
\sin{\theta} & \cos{\theta} & y \\
0 & 0 & 1
\end{pmatrix} \begin{pmatrix} a \\ b \\ 1 \end{pmatrix} = \begin{pmatrix}
a \cos{\theta} - b \sin{\theta} + x \\
a \sin{\theta} + b \cos{\theta} + y \\
1 \end{pmatrix} = \textbf{v}'
\end{equation*}
so that we obtain a new vector $\textbf{v}'$ in $\mathbb{R}^2$.

The three-dimensional Lie algebra of $E_2$, denoted $\mathfrak{e}_2$, is defined by the commutation relations
\begin{equation}\label{e_2 commutation rel}
[X,Y]=0, \,\,\,\, [Z,X]=Y, \,\,\,\,[Y,Z]=X.
\end{equation}
where $X, Y, Z$ are basis operators.  A $3 \times 3$ matrix representation is given by taking 
\begin{equation*}
A=\begin{pmatrix}
0 & 0 & 1 \\
0 & 0 & 0 \\
0 & 0 & 0 
\end{pmatrix} , \,\,\,\, B = 
\begin{pmatrix}
0 & 0 & 0 \\
0 & 0 & 1 \\
0 & 0 & 0 
\end{pmatrix} , \,\,\,\, C = 
\begin{pmatrix}
0 & -1 & 0 \\
1 & 0 & 0 \\
0 & 0 & 0 
\end{pmatrix} 
\end{equation*}
with the association $X \rightarrow A, Y \rightarrow B, Z \rightarrow C$.  This representation, however, is not unitary.
In order to derive the Bessel functions and their properties, it is necessary to use a unitary differential operator representation of $\mathfrak{e}_2$.  One example of this is given by defining\footnote{This definition assumes the continuous basis $|{\bf x}\rangle=|x,y\rangle$.}
\begin{equation}\label{e_2 P_x P_y L_z}
P_x = \frac{\partial}{\partial x}, \,\,\,\, P_y = \frac{\partial}{\partial y}, \,\,\,\, L_z = x \frac{\partial}{\partial y} - y \frac{\partial}{\partial x}.
\end{equation}  
These operators satisfy the same commutation relations given in (\ref{e_2 commutation rel}) if we make the identification $X \rightarrow P_x$, $Y \rightarrow(-P_y)$, $Z \rightarrow L_z$ since 
\begin{equation}\label{e_2 rep comm rel}
[P_x,-P_y]=0, \,\,\,\, [L_z,P_x]=-P_y, \,\,\,\,[-P_y,L_z]=P_x.
\end{equation}  Hence, these differential operators also work as a representation for $\mathfrak{e}_2$.  

\section{The Bessel Functions as Basis Functions of the Representation Space}

The operators $P_x$, $P_y$ and $L_z$ act on a vector space $V$ consisting of continuous ($C^\infty$) functions of two variables.  This vector space can be described using both a continuous basis and discrete basis, with basis functions denoted $|x,y\rangle$ and $|n\rangle$, respectively.  The continuous basis can also be expressed in polar coordinate form, written $|r,\phi\rangle$.  The Bessel functions are proportional to the mixed basis functions $\langle r,\phi | n \rangle$ of the representation space $V$.  Note that the mixed basis functions can also be viewed as matrix elements $\langle r,\phi | I | n \rangle$ of the identity operator.    

To compute the mixed basis functions, it is very convenient to first define the operators $P_+ = iP_x - P_y$ and $P_- = iP_x + P_y$.  
They are useful since they act as ladder operators on the set of basis functions of $V$.  That is, if $| n \rangle$ is a basis function of $V$, then (see \cite{Miller book})
\begin{equation} \label{e_2 ladder ops}
P_+ | n \rangle = -|n+1 \rangle ,\,\,\,\, P_-| n \rangle = -|n-1 \rangle .
\end{equation}
It is possible to choose the basis functions $| n \rangle$ to be eigenfunctions of $L_z$ with eigenvalue $n$:
\begin{equation} \label{L_z eig eq}
L_z | n \rangle = n | n \rangle
\end{equation}
We will exploit these relationships in order to derive the Bessel functions and their properties.  It is helpful to express the operators $P_+, P_-, L_z$ in polar coordinate form:
\begin{equation}\label{e_2 diff op rep}
L_z = -i \frac{\partial}{\partial \phi}, \,\,\,\, P_{\pm} = e^{\pm i\phi} \left( \pm \frac{\partial}{\partial r} + \frac{i}{r}\frac{\partial}{\partial \phi} \right).
\end{equation}
 
Using (\ref{L_z eig eq}), we can solve for the general form satisfied by the mixed basis functions:
\begin{align*}
-i \frac{\partial \langle r,\phi | n \rangle}{\partial \phi} &= n \langle r,\phi | n \rangle \\
\Rightarrow \,\,\,\, \langle r,\phi | n \rangle &= J_n(r) \,e^{in\phi},
\end{align*}
for some function $J_n(r)$.   It will be shown in the next section that the functions $J_n(r)$ are the familiar Bessel functions. 

\section{Differential Equation}\label{sec:bessel diff eq}

It is possible to obtain the differential equation (\ref{eq:bessel diff eq}) satisfied by the Bessel functions by computing $P_+P_- \langle r, \phi | n \rangle$\footnote{We are actually computing $\langle r,\phi | P_+P_-  | n \rangle$.  Although writing $P_+P_-\langle r,\phi | n \rangle$ is technically undefined, it keeps the calculation more concise.  For more details, see Appendix \ref{app:bessel extra details}.}.   We compute this expression using both the discrete and the continuous basis, similar to the procedure used for the Hermite polynomials.  

First, we consider the discrete basis $| n \rangle$. Since the raising and lowering angular momentum operators satisfy (\ref{e_2 ladder ops}), 
we have
\begin{equation} \label{bessel diff discrete}
P_+P_-\langle r, \phi | n \rangle = P_+ (-\langle r, \phi | n-1 \rangle) = \langle r, \phi | n \rangle.
\end{equation}
In other words, $\langle r,\phi | n \rangle$ is an eigenfunction of $P_+P_-$ with an eigenvalue equal to $1$.

For the continuous basis $|r,\phi \rangle$, we compute $P_+P_-$ using (\ref{e_2 diff op rep}):
\begin{align*}
P_+P_- \langle r, \phi | n \rangle &= P_+e^{-i\phi} \left(-\frac{\partial}{\partial r} + \frac{i}{r} \frac{\partial}{\partial \phi} \right) J_n(r) e^{in\phi} \\
&=P_+ \left(-J_n'(r) - \frac{n}{r}J_n(r)  \right) e^{i(n-1)\phi} \\
&= e^{i\phi}\left( \frac{\partial}{\partial r} + \frac{i}{r} \frac{\partial}{\partial \phi}\right) \left(-J_n'(r) - \frac{n}{r}J_n(r)  \right) e^{i(n-1)\phi} .
\end{align*}
After taking derivatives and doing the algebra you get
\begin{equation}\label{bessel diff continuous}
P_+P_- \langle r, \phi | n \rangle = \left(-J_n''(r) - \frac{1}{r}J_n'(r) + \frac{n^2}{r^2}J_n(r)\right) e^{in\phi}.
\end{equation}
Equating (\ref{bessel diff discrete}) and (\ref{bessel diff continuous}) and canceling out $e^{in\phi}$ from both sides gives Bessel's equation:
\begin{equation*}
J_n''(r) + \frac{1}{r}J_n'(r)+\left(1-\frac{n^2}{r^2}\right)J_n(r) = 0.
\end{equation*}
Hence, the functions $J_n(r)$ are the familiar Bessel functions.

\section{Differential and Recursion Relations}\label{sec:bessel rec diff}

The recursion relation (\ref{eq:bessel recursion}) for the Bessel functions is derived by computing $(P_+ + P_-) \langle r, \phi | n \rangle$.  By (\ref{e_2 ladder ops}), we can write
\begin{align*}
P_+ \langle r, \phi | n \rangle &= -\langle r, \phi | n+1 \rangle.
\end{align*}
Then, using the fact that $\langle r, \phi | n \rangle = J_n(r) e^{in\phi}$ and writing $P_+$ according to (\ref{e_2 diff op rep}), we get:
\begin{align}
e^{i\phi} \left( \frac{\partial}{\partial r} + \frac{i}{r}\frac{\partial}{\partial \phi} \right) J_n(r)e^{in\phi} &= -J_{n+1}(r) e^{i(n+1)\phi} \notag \\
\Rightarrow \;\;\;\; J_n'(r) -\frac{n}{r}J_n(r) &= -J_{n+1}(r) \label{bessel diff rel. 1}
\end{align}
after dividing out the exponential terms.  Similarly, using 
\begin{equation*}
P_- \langle r, \phi | n \rangle = -\langle r, \phi | n-1 \rangle
\end{equation*} it is easy to show that
\begin{equation}\label{bessel diff rel. 2}
-J_n'(r)-\frac{n}{r}J_n(r) = -J_{n-1}(r).
\end{equation}
Adding (\ref{bessel diff rel. 1}) and (\ref{bessel diff rel. 2}) gives the standard recursion relation for the Bessel functions:
\begin{equation*}
\frac{2n}{r}J_n(r) = J_{n-1}(r)+J_{n+1}(r).
\end{equation*}
The differential relation (\ref{eq:bessel diff rel}) is derived by computing $(P_+-P_-)\langle r,\phi | n \rangle$.  For this computation, just subtract (\ref{bessel diff rel. 2}) from (\ref{bessel diff rel. 1}).  You get:
\begin{equation*}
2J_n'(r)=J_{n-1}(r)-J_{n+1}(r).
\end{equation*}

\section{Generating Function}\label{sec:bessel generating function}

One particular generating function satisfied by the Bessel functions can be derived by computing the matrix elements of the exponential $e^{tP_+}$ in the mixed basis.  The exponential of the $P_+$ operator is an element of the Euclidean group ($E_2$) since $P_+$ lives in the Euclidean algebra ($\mathfrak{e}_2$).  To derive the generating function, we compute the matrix elements in two different ways and equate the results. 

First, expand the exponential operator in a Taylor series.  For each term in the series, $P_+$ operates on the basis functions $\langle r,\phi | n \rangle$, so that
\begin{align}
\langle r, \phi | e^{tP_+} | n \rangle &= \sum_{m=0}^\infty \frac{t^m P_+^m}{m!} \langle r,\phi | n \rangle \notag \\
&= \sum_{m=0}^\infty \frac{(-1)^m t^m}{m!} e^{i(n+m)\phi} J_{n+m}(r). \label{bessel gen func first step}
\end{align}

Alternatively, we can write $P_+$ in terms of the operators $P_x = \frac{\partial}{\partial x}$ and $P_y = \frac{\partial}{\partial y}$ and exploit the fact that these operators act as shift operators.  We saw in the derivation of the Hermite generating function that $\frac{\partial}{\partial x}$ shifts the $x$ argument of a function $f(x)$ since
\begin{equation*}
e^{-t \frac{\partial}{\partial x}} f(x) = f(x-t).
\end{equation*}
A simple way to see why these operators shift the argument of a point in the Euclidean plane is best illustrated by writing them using the $3 \times 3$ matrix representation:
\begin{equation*}
P_x=\begin{pmatrix}
0 & 0 & 1 \\
0 & 0 & 0 \\
0 & 0 & 0 
\end{pmatrix} , \,\,\,\, P_y = 
\begin{pmatrix}
0 & 0 & 0 \\
0 & 0 & 1 \\
0 & 0 & 0 
\end{pmatrix}.
\end{equation*}
Since these operators are \textit{nilpotent} matrices of degree two, i.e., since these operators satisfy $P_x^2 = P_y^2 = 0$, it is easy to compute the corresponding Euclidean group operators via a power series expansion:
\begin{equation*}
e^{tP_x} = I + tP_x = \begin{pmatrix}
1 & 0 & t \\
0 & 1 & 0 \\
0 & 0 & 1 
\end{pmatrix},
\end{equation*}
\begin{equation*}
e^{tP_y} = I + tP_y = \begin{pmatrix}
1 & 0 & 0 \\
0 & 1 & t \\
0 & 0 & 1 
\end{pmatrix}.
\end{equation*}
These group elements operate on points in the plane by shifting their $x$ and $y$ coordinates by an amount $t$:
\begin{equation*}
\begin{pmatrix}
1 & 0 & t \\
0 & 1 & 0 \\
0 & 0 & 1 
\end{pmatrix}
\begin{pmatrix}
x \\
y \\
1
\end{pmatrix} = \begin{pmatrix}
x+t \\
y \\
1
\end{pmatrix}, 
\end{equation*}
\begin{equation*}
\begin{pmatrix}
1 & 0 & 0 \\
0 & 1 & t \\
0 & 0 & 1 
\end{pmatrix}
\begin{pmatrix}
x \\
y \\
1
\end{pmatrix} = \begin{pmatrix}
x\\
y+t \\
1
\end{pmatrix}.
\end{equation*}

To exploit this fact, we write $P_+$ in terms of $P_x$ and $P_y$ and make the substitution $r = \sqrt{x^2 + y^2}$.  Then we have,
\begin{equation} \label{eq:bessel gen func cont}
\begin{aligned}
\langle r,\phi | e^{tP_+} | n \rangle &= e^{t(iP_x-P_y)}\langle r,\phi | n \rangle \\
&= e^{in\phi}e^{itP_x}e^{-tP_y}J_n(\sqrt{x^2+y^2})\\
&= e^{in\phi}e^{itP_x}J_n(\sqrt{x^2+(y-t)^2}) \\
&= e^{in\phi}J_n(\sqrt{(x+it)^2+(y-t)^2})
\end{aligned}
\end{equation}
Unlike what was the case for the derivation of the Hermite generating function, no disentangling theorem is needed here because $P_x$ and $P_y$ commute ($[P_x,P_y]=0$).  This last expression can be cleaned up a bit by expanding the terms inside the square root.  This gives:
\begin{equation}\label{bessel gen func second step}
\langle r,\phi | e^{tP_+} | n \rangle = e^{in\phi}J_n(\sqrt{r^2+2t(ix-y)}).
\end{equation}

Equating (\ref{bessel gen func first step}) and (\ref{bessel gen func second step}) and canceling the term $e^{in\phi}$ from both sides gives the Bessel generating function (\ref{eq:bessel gen func 1}):
\begin{equation*}
J_n(\sqrt{r^2+2t(ix-y)}) = \sum_{m=0}^\infty \frac{(-1)^m t^m}{m!} e^{im\phi} J_{n+m}(r).
\end{equation*}
In Appendix \ref{app:another bessel gen func}, we use another method to derive a different generating function.

\section{Group Contraction}\label{e_2 group contraction}

The operators $P_x, P_y, P_\pm,$ and $L_z$ as defined in (\ref{e_2 P_x P_y L_z}) and (\ref{e_2 diff op rep}) should look familiar.  They are closely related to the angular momentum operators $L_x, L_y, L_\pm,$ and $L_z$ defined by
\begin{equation*}
L_x = y \frac{\partial}{\partial z} - z \frac{\partial}{\partial y}, \,\,\,\, L_y = z \frac{\partial}{\partial x}-x \frac{\partial}{\partial z}, \,\,\,\,L_z = x \frac{\partial}{\partial y} - y \frac{\partial}{\partial x},
\end{equation*}
\begin{equation*}
L_+ = L_x + i L_y,\,\,\,\,L_- = L_x - iL_y.
\end{equation*}
In spherical coordinates, $L_z, L_\pm$ can be written (cf.\ \cite{Griffiths book})
\begin{equation}\label{so(3) diff op rep}
L_z = -i \frac{\partial}{\partial \phi}, \,\,\,\, L_{\pm} = e^{\pm i\phi} \left( \frac{\partial}{\partial \theta} \pm i \cot{\theta} \frac{\partial}{\partial \phi} \right).
\end{equation}

The operators $L_x, L_y, L_z$ form a representation of the Lie algebra $\mathfrak{so}(3)$, the algebra of the three-dimensional rotation group $SO(3)$, and act on a vector space $V$ consisting of continuous ($C^\infty$) functions of three variables that lie on a two-dimensional manifold $x^2+y^2+z^2 = $ constant.  They satisfy the commutation relations
\begin{equation}\label{so(3) comm rel}
[L_x, L_y] = -L_z, \,\,\,\, [L_y, L_z ] = -L_x, \,\,\,\, [L_z, L_x] = -L_y.
\end{equation}
The fact that these operators, together with the ladder operators $L_\pm$, look remarkably similar to their $\mathfrak{e}_2$ counterparts  is no accident.  It turns out that the groups $E_2$ and $SO(3)$ are related by a procedure called \textit{group contraction}.  Group contraction is the process of obtaining one group from another by applying some kind of limiting operation to the latter.  

By applying a limiting operation to $SO(3)$, the Euclidean group $E_2$ is obtained.  This relationship provides an alternative way to derive the the Bessel functions, different from the method used in Section \ref{sec:bessel diff eq}.  Just as the Hermite polynomials are related to the Heisenberg group and the Bessel functions are related to the Euclidean group in the plane, it can be shown that the associated Legendre polynomials $P_m^l(\cos{\theta})$ are related to the three-dimensional rotation group.  Since $E_2$ is the contracted limit of $SO(3)$, the Bessel functions can be derived by taking suitable limits of the associated Legendre polynomials.

\subsection{Computing the limit $SO(3)\rightarrow E_2$} 

The specific limiting operation applied to $SO(3)$ which we will use requires us to construct a continuous sequence of basis transformations.  The basis vectors $L_x, L_y, L_z$ in our representation of $\mathfrak{so}(3)$ are, of course, not unique.  We can define a new set of basis vectors $L_x', L_y', L_z'$ by the transformation
\begin{equation}\label{so(3) basis transformation}
\begin{pmatrix}
L_x' \\
L_y' \\
L_z'
\end{pmatrix} = \begin{pmatrix}
1/R  & 0 & 0 \\
0 & 1/R & 0 \\
0 & 0 & 1 
\end{pmatrix} \begin{pmatrix}
L_x \\
L_y \\
L_z
\end{pmatrix}.
\end{equation}
This transformation defines a new basis in $\mathfrak{so}(3)$ since $L_x', L_y', L_z'$ satisfy the same commutation relations (cf.\ (\ref{so(3) comm rel})) as $L_x, L_y, L_z$ (up to a constant factor):
\begin{equation}\label{basis trans comm rel}
[L_x', L_y'] = \frac{-L_z'}{R^2},\,\,\,\,[L_y',L_z'] = -L_x',\,\,\,\,[L_z',L_x'] = -L_y'.
\end{equation}
Note that for $0 < R < \infty$ the the transformation matrix in (\ref{so(3) basis transformation}) is invertible.  This is necessary in order that the Lie algebra $\mathfrak{so}(3)$ be unchanged \cite{Gilmore book}.  Note, however, that if we take the limit of (\ref{so(3) basis transformation}) as $R \rightarrow \infty$, the transformation becomes singular:
\begin{equation*}
 \begin{pmatrix}
1/R  & 0 & 0 \\
0 & 1/R & 0 \\
0 & 0 & 1 
\end{pmatrix} \rightarrow 
 \begin{pmatrix}
0  & 0 & 0 \\
0 & 0 & 0 \\
0 & 0 & 1 
\end{pmatrix}.
\end{equation*}
Hence, the change of basis is no longer well defined.  In the limit as $R \rightarrow \infty$ the commutation relations given in (\ref{basis trans comm rel}), however, remain well defined.  In fact, we can show that 
\begin{equation}\label{so(3) limits}
\lim_{R\rightarrow \infty}{L_x'} =- P_y, \,\,\,\, \lim_{R\rightarrow \infty}{L_y'} = P_x,
\end{equation}
where $P_x$ and $P_y$ are the basis operators of the unitary differential operator representation of $\mathfrak{e}_2$ defined in (\ref{e_2 P_x P_y L_z}).  Then the commutation relations in (\ref{basis trans comm rel}) become
\begin{equation*}
[-P_y,P_x]=0,\,\,\,\,[P_x,L_z]=P_y,\,\,\,\,[L_z,-P_y]=-P_x,
\end{equation*}
which are the familiar commutation relations satisfied by the basis operators of $\mathfrak{e}_2$ and are equivalent to (\ref{e_2 rep comm rel}).  
This shows that $E_2$ is the contracted limit of $SO(3)$.  Note that $L_z' = L_z$.

Proving that the limits in (\ref{so(3) limits}) converge to the basis operators of $\mathfrak{e}_2$ is relatively simple if without loss of generality (cf.\ \cite{Gilmore book}) we let the operators in $\mathfrak{so}(3)$ act in the neighborhood of the point $(0,0,R)$ expressed in cartesian coordinates and lying on a sphere of radius $R$ centered at the origin.  Calculating the limits at this point makes it easy to show that several terms in the limit vanish.  For example,
\begin{align*}
\lim_{R\rightarrow \infty}{L_x'} = \lim_{R\rightarrow \infty}{\frac{L_x}{R}} &= \lim_{R\rightarrow \infty}{\frac{1}{R} \left(  y \frac{\partial}{\partial z} - z \frac{\partial}{\partial y}  \right)} \\
&= \lim_{R\rightarrow \infty}{ \frac{1}{R} \left( y \frac{\partial}{\partial z}-R \frac{\partial}{\partial y} \right) } \\
&= -\frac{\partial }{\partial y} \\
&= -P_y.
\end{align*}
Proving the other limit in (\ref{so(3) limits}) is similar.

It is also easy to show that the ladder operators in $\mathfrak{so}(3)$ converge to the ladder operators in $\mathfrak{e}_2$.  We will illustrate this fact in two different ways.  First, observe that we can directly compute
\begin{align*}
\lim_{R\rightarrow \infty}\frac{L_\pm}{R} &= \lim_{R\rightarrow \infty} \frac{L_x}{R}\pm \frac{iL_y}{R} \\
&= \lim_{R\rightarrow \infty} L_x' \pm iL_y' \\
&= -P_y \pm iP_x \\
&= \pm P_\pm.
\end{align*}

Alternatively, it is useful to compute the limit using polar coordinates in order to illustrate exactly how this works geometrically.  The rotation group $SO(3)$ can be contracted to $E_2$ for the same reason that the earth, a sphere, appears flat over a small enough region on the surface.  To see this, suppose that the rotation group operates on a point $\textbf{p}$ in $\mathbb{R}^3$.  For convenience, choose a coordinate system so that $\textbf{p} = (R,0,0)$ in spherical coordinates.   Clearly, $\textbf{p}$ lies at the north pole of a sphere of radius $R$ centered at the origin.  Consider a point $\textbf{q}$ on the tangent plane containing $\textbf{p}$ and lying perpendicular to the z-axis.   Suppose $\textbf{q}$ has coordinates $(s,\theta,\phi)$ and lies at a distance $r$ from $\textbf{p}$.  Then $\sin{\theta} = r/s$ and $\cos{\theta} = R/s$ so that 
\begin{equation}\label{cotangent}
\cot{\theta} = \frac{R}{r}.
\end{equation}

We now take the limit of the angular momentum ladder operators $L_{\pm} / R$ as $R$ goes to infinity:
\begin{equation*}
\lim_{R\rightarrow \infty} \frac{L_\pm}{R} = e^{\pm i\phi} \lim_{R\rightarrow \infty} \left( \frac{1}{R} \frac{\partial}{\partial \theta} \pm \frac{i \cot{\theta}}{R} \frac{\partial}{\partial \phi} \right).
\end{equation*}
The second term in parentheses can be simplified using (\ref{cotangent}) to cancel out the $R$ in the denominator.  For the first term, we can write
\begin{align*}
\frac{\partial}{\partial \theta} = \frac{\partial r}{\partial \theta} \frac{\partial}{\partial r} + \frac{\partial s}{\partial \theta} \frac{\partial}{\partial s} 
\end{align*}
by the multivariate chain rule.  But $r = s \sin{\theta}$ implies 
\begin{equation*}
\frac{\partial r}{\partial \theta} = s\cos{\theta} = R.
\end{equation*}
Similarly, $s = R \sec{\theta}$ implies 
\begin{equation*}
\frac{\partial s}{\partial \theta} = R \tan{\theta}\sec{\theta} = r \sec{\theta}.
\end{equation*}
Plugging it all in, we get
\begin{equation*}
\frac{\partial}{\partial \theta} = R \frac{\partial}{\partial r} +  r \sec{\theta} \frac{\partial}{\partial s}
\end{equation*}
and so
\begin{align*}
\lim_{R\rightarrow \infty} \frac{L_\pm}{R} &= e^{\pm i\phi}\lim_{R\rightarrow \infty} \left( \frac{1}{R}R \frac{\partial}{\partial r} + \frac{r\sec{\theta}}{R} \frac{\partial}{\partial s} \pm \frac{i}{R} \frac{R}{r}\frac{\partial}{\partial \phi} \right) \\
&= e^{\pm i\phi} \left(  \frac{\partial}{\partial r} \pm \frac{i}{r}\frac{\partial}{\partial \phi} \right) \\
&=\pm P_{\pm}.
\end{align*}

\subsection{Computing the limit $P_m^l(\cos{\theta}) \rightarrow J_m(r)$}

The most straightforward way to show that the associated Legendre polynomials converge to the Bessel functions is to show that the Legendre equation converges to the Bessel equation in the appropriate limit.  The associated Legendre polynomials $P_m^l(z)$ are defined to be the solution of the Legendre equation:
\begin{equation*}
\left[\frac{d}{dz}(1-z^2)\frac{d}{dz} + l(l+1) - \frac{m^2}{1-z^2}\right]P_m^l(z)=0.
\end{equation*}
The Bessel equation is obtained by setting $z=\cos{\theta}$ and taking the limit of the Legendre equation as $\theta \rightarrow 0$ and $l \rightarrow \infty$.  The limit is taken with $l \theta = r$.  

To illustrate, first let $z=\cos{\theta}$.  Then the Legendre equation becomes
\begin{equation*}
\left[\frac{d}{d(\cos{\theta})}(1-\cos^2{\theta})\frac{d}{d(\cos{\theta})} + l(l+1) - \frac{m^2}{1-\cos^2{\theta}}\right]P_m^l(\cos{\theta})=0.
\end{equation*}
But since
\begin{equation*}
\frac{d}{d(\cos{\theta})} = \frac{d\theta}{d(\cos{\theta})}\frac{d}{d\theta}=-\frac{1}{\sin{\theta}} \frac{d}{d\theta},
\end{equation*}
and using a standard trig identity, we can rewrite the Legendre equation as
\begin{equation*}
\left[\frac{1}{\sin{\theta}} \frac{d}{d\theta} \sin{\theta} \frac{d}{d\theta} + l(l+1) - \frac{m^2}{\sin^2{\theta}}\right]P_m^l = 0.
\end{equation*}

Next, take the limit of this equation as $\theta \rightarrow 0$, $l \rightarrow \infty$, and $l\theta = r$.  This gives
\begin{equation*}
\left[\frac{1}{\theta} \frac{d}{d\theta} \theta \frac{d}{d\theta} + l^2 - \frac{m^2}{\theta^2}\right]J_m = 0,
\end{equation*}
or equivalently,
\begin{equation*}
\left[\theta \frac{d}{d\theta} \theta \frac{d}{d\theta} + r^2 - m^2\right]J_m = 0
\end{equation*}
after multiplying both sides by $\theta^2$.  Here, $J_m$ is some function of $r$.  If we make the substitution $\theta = r/l$ we can write
\begin{equation*}
\left[\frac{r}{l} \frac{d}{d(r/l)} \frac{r}{l} \frac{d}{d(r/l)} + r^2 - m^2\right]J_m = 0.
\end{equation*}
But since
\begin{equation*}
\frac{d}{d(r/l)} = \frac{dr}{d(r/l)}\frac{d}{dr} = l \frac{d}{dr},
\end{equation*}
we get that
\begin{equation*}
\left[r\frac{d}{dr}r\frac{d}{dr}+r^2-m^2\right]J_m=0.
\end{equation*}
This is, in fact, the Bessel equation.  Dividing both sides by $r^2$ and expanding the derivatives using the product rule gives the form of the Bessel equation given earlier:
\begin{equation*}
\left[\frac{d^2}{dr^2}+\frac{1}{r}\frac{d}{dr}+1-\frac{m^2}{r^2}\right]J_m(r) = 0.
\end{equation*}




\chapter{Summary}

In this thesis, we have explored the relationship between some of the special functions in mathematical physics and the theory of Lie groups and Lie algebras.  Although we focused specifically on the Hermite polynomials and Bessel functions, the methods outlined in this thesis can be applied to many other kinds of special functions.

Our general procedure is summarized in Figure \ref{fig:big picture}.  We began by representing the elements of a Lie group and Lie algebra as linear operators acting on a space of continuous functions.  We saw that the operators in the Lie algebra can be expressed as infinite-dimensional matrices in the discrete and continuous bases of the representation space.  In the discrete basis, the operators took the form of infinite-dimensional matrices with one nonzero entry.  These operators acted as ladder operators on the basis functions of the representation space.  In the continuous basis, the operators took the form of first-order differential operators.  

We then showed that the special functions are proportional to the mixed basis functions in the representation space.  By computing the matrix elements of linear combinations of operators in the mixed basis, we were able to derive properties such as differential and recursion relations, differential equations, and generating functions.  To compute the matrix elements, we performed a resolution of the identity in two different ways (using both bases) and equated the results.  

We concluded with a brief discussion of group contraction, which demonstrated that special functions can also be obtained by taking suitable limits.

Our contribution in this thesis was to present existing results found in the literature in a way that can be understood by undergraduate physics students who have studied quantum mechanics.   For this reason, we made use of bra-ket notation extensively in order to derive the results.  A physics student who is exposed to the ideas presented in this thesis will never look at the quantum harmonic oscillator problem or quantum mechanics the same way again.


\appendix	 

\chapter{Properties of the Special Functions}\label{app:table}

The properties that are derived in this thesis are listed below.  Note that the Hermite polynomials and Bessel functions satisfy \textit{many} additional properties besides what is given here. 

\section{Hermite Polynomials}\label{sec:hermite table}

Rodrigues's formula \cite{Gilmore book}:
\begin{equation}\label{hermite rod}
H_n(x) = e^{x^2} \left(-\frac{d}{dx}\right)^n e^{-x^2} = e^{x^2/2} \left(x-\frac{d}{dx}\right)^n e^{-x^2/2}
\end{equation}
Differential equation \cite{Vilenkin book}:
\begin{equation}\label{hermite diff eq}
H_n''(x)-2xH_n'(x)+2nH_n(x)=0
\end{equation}
Recursion relation \cite{Vilenkin book}
\begin{equation}\label{hermite recursion}
H_{n+1}(x) -2xH_n(x)+2nH_{n-1}(x)=0
\end{equation}
Differential relation \cite{Vilenkin book}:
\begin{equation}\label{hermite diff rel}
H_n'(x)=2nH_{n-1}(x)
\end{equation}
Generating function \cite{Vilenkin book}:
\begin{equation}\label{hermite gen func}
e^{2xt-t^2}= \sum_{n=0}^\infty \frac{H_n(x)}{n!} t^n.
\end{equation}

\section{Bessel Functions}\label{sec:bessel table}

Differential equation:
\begin{equation}\label{eq:bessel diff eq}
J_n''(r) + \frac{1}{r}J_n'(r)+\left(1-\frac{n^2}{r^2}\right)J_n(r) = 0.
\end{equation}
Recursion relation:
\begin{equation}\label{eq:bessel recursion}
\frac{2n}{r}J_n(r) = J_{n-1}(r)+J_{n+1}(r).
\end{equation}
Differential relations:
\begin{align}
J_n'(r) -\frac{n}{r}J_n(r) &= -J_{n+1}(r) \\
-J_n'(r)-\frac{n}{r}J_n(r) &= -J_{n-1}(r) \\
2J_n'(r)&=J_{n-1}(r)-J_{n+1}(r) \label{eq:bessel diff rel}
\end{align}
Generating functions:
\begin{align}
e^{in\phi}J_n(\sqrt{r^2+2t(ix-y)}) &= \sum_{m=0}^\infty \frac{(-1)^m t^m}{m!} e^{i(n+m)\phi} J_{n+m}(r) \label{eq:bessel gen func 1} \\
\exp\left( \frac{r\phi}{\sqrt{2r\phi t + r^2 }}\right) J_n\left(\sqrt{2r\phi t + r^2 }\right) &= \sum_{m=0}^\infty \frac{(-1)^m t^m}{m!} e^{i(n+m)\phi} J_{n+m}(r)  \label{eq:bessel gen func 2}
\end{align}

\chapter{Additional Details for Chapter \ref{ch:bessel functions}}\label{app:bessel extra details}
In Chapter \ref{ch:hermite polynomials}, we derived the properties of the Hermite polynomials by computing the matrix elements in the mixed basis of different linear combinations of operators in the Heisenberg group and algebra.  In Chapter \ref{ch:bessel functions}, we followed this same procedure using the Euclidean group in the plane.  However, the calculations in Chapter \ref{ch:bessel functions} left out some of the details in an effort to be concise.  In this appendix, we fill in the missing details.  The following equations are needed:

\begin{align*}
\langle r',\phi' | r,\phi \rangle &= \delta\left( \frac{1}{2}r'^2  - \frac{1}{2}r^2\right)\delta(\phi'-\phi) \\
\langle r',\phi' | P_\pm | r,\phi \rangle &=e^{\pm i\phi} \left( \pm \frac{\partial}{\partial r} + \frac{i}{r}\frac{\partial}{\partial \phi} \right)\delta\left( \frac{1}{2}r'^2 - \frac{1}{2}r^2\right)\delta(\phi'-\phi) \\
\langle n' | n \rangle &= \delta_{n',n} \\
\langle n' | P_+ | n \rangle &= -\delta_{n',n+1} \\
\langle n' | P_- | n \rangle &= -\delta_{n',n-1}
\end{align*}

\section{Matrix Elements of $P_\pm$ in Mixed Basis}

In section \ref{sec:bessel diff eq} we derived the Bessel equation by computing $P_+P_-\langle r,\phi | n \rangle$.  In section \ref{sec:bessel rec diff}, we derived the recursion and differential relations by computing $(P_++P_-)\langle r,\phi | n \rangle$ and $(P_+-P_-)\langle r,\phi | n \rangle$, respectively. This was a sloppy use of notation since $P_+$ and $P_-$ are not well defined until  we project them into a particular basis.  
For this reason, we must actually compute the matrix elements of the operators in the mixed basis.  We illustrate this procedure for the $P_+$ operator:

\begin{align*}
\langle r,\phi | P_+ | n \rangle &= \int \langle r,\phi | P_+| r',\phi' \rangle \langle r',\phi' | n \rangle\, r'dr'd\phi' \\
&= e^{ i\phi} \left(  \frac{\partial}{\partial r} + \frac{i}{r}\frac{\partial}{\partial \phi} \right)\int \delta\left( \frac{1}{2}r^2 - \frac{1}{2}r'^2\right)\delta(\phi-\phi')  \langle r',\phi' | n \rangle\, r'dr'd\phi' \\
&= e^{ i\phi} \left(  \frac{\partial}{\partial r} + \frac{i}{r}\frac{\partial}{\partial \phi} \right) \int \left[ \frac{\delta(r-r')}{r'} + \frac{\delta(r+r')}{r'}\right]\langle r',\phi | n \rangle \, r'  dr'\\
&= e^{ i\phi} \left(  \frac{\partial}{\partial r} + \frac{i}{r}\frac{\partial}{\partial \phi} \right) \langle r,\phi | n \rangle
\end{align*}

In the spirit of (\ref{eq:standard procedure}), it is also necessary to perform a resolution identity for the discrete basis, but that is more straightforward and we omit it.

\section{Matrix Elements of $e^{tP_+}$ in Mixed Basis}
In section \ref{sec:bessel generating function}, we made the assumption that $\langle r,\phi | e^{tP_+}|n\rangle = e^{tP_+}\langle r,\phi | n \rangle$.  In this section, we compute the matrix elements more rigorously.  Again, the reason this is necessary is because the group element $e^{tP_+}$ is not well defined until we project it into a particular basis. 

First, we project the exponential into the discrete basis:
\begin{align}
\langle r,\phi | e^{tP_+} | n \rangle &= \sum_{n'=0}^\infty \langle r,\phi | n' \rangle \langle n' | e^{tP_+} | n \rangle \notag \\
&= \sum_{n'=0}^\infty \langle r,\phi | n' \rangle \langle n' | \sum_{m=0}^\infty \frac{t^m P_+^m}{m!} | n \rangle \notag \\
&= \sum_{n'=0}^\infty \langle r,\phi | n' \rangle \langle n' | \sum_{m=0}^\infty \frac{(-1)^m t^m}{m!} | n+m \rangle \notag \\ 
&= \sum_{n'=0}^\infty \langle r,\phi | n' \rangle \sum_{m=0}^\infty \frac{(-1)^m t^m}{m!} \langle n' | n+m \rangle \notag\\ 
&= \sum_{n'=0}^\infty \langle r,\phi | n' \rangle \sum_{m=0}^\infty \frac{(-1)^m t^m}{m!} \delta_{n',n+m} \label{includes delta}
\end{align}
since $\langle n' | n+m \rangle = \delta_{n',n+m}$.  For a fixed $m$, the expression is 0 unless $n' = n+m \geq 0$.  Hence (\ref{includes delta}) becomes
\begin{align*}
\langle r,\phi | e^{tP_+} | n \rangle &=\sum_{m=0}^\infty \langle r,\phi | n+m \rangle \frac{(-1)^{m} t^{m}}{m!} \\
&= \sum_{m=0}^\infty \frac{(-1)^m t^m}{m!} e^{i(n+m)\phi} J_{n+m}(r)
\end{align*}
which is equivalent to (\ref{bessel gen func first step}).

Next, we project the exponential into the continuous basis.  Using the fact that (cf. \cite{Gilmore gen func})
\begin{equation*}
\langle r,\phi | e^{tP_+} | r',\phi' \rangle = e^{t(iP_x-P_y)}\delta\left( \frac{1}{2}r^2 - \frac{1}{2}r'^2\right)\delta(\phi-\phi'),
\end{equation*}
we can write
\begin{align*}
\langle r,\phi | e^{tP_+} | n \rangle &= \int \langle r,\phi | e^{tP_+} | r',\phi' \rangle \langle r',\phi' | n \rangle \, r' dr' d\phi' \\
&= e^{t(iP_x-P_y)} \int \delta\left( \frac{1}{2}r^2 - \frac{1}{2}r'^2\right)\delta(\phi-\phi') \langle r',\phi' | n \rangle \, r' dr' d\phi' \\
&= e^{t(iP_x-P_y)} \int  \left[ \frac{\delta(r-r')}{r'} + \frac{\delta(r+r')}{r'}\right] \langle r',\phi | n \rangle \, r' dr' \\
&= e^{t(iP_x-P_y)}\langle r,\phi | n \rangle.
\end{align*}
We can then proceed as in (\ref{eq:bessel gen func cont}).

\chapter{Another Bessel Generating Function}\label{app:another bessel gen func}
There are many different examples of Bessel generating functions beside the one computed in  Chapter \ref{ch:bessel functions}.  In this appendix we derive another generating function.  The derivation given here is based off of the one found in \cite{Miller article}.  

To begin, we will assume that the action of the group element $e^{tP_+}$ on the mixed basis functions $\langle r_0,\phi_0 | n \rangle =: f_n(r_0,\phi_0)$ shifts the arguments of the functions and scales them by a factor of $q$.  We will also assume that the shifts in the arguments and the scaling factor are functions of a real number $t$ with initial conditions $q(0) = 1, r(0) = r_0, \phi(0) = \phi_0$.  Then,
\begin{align}\label{eq:second gen func initial}
e^{tP_+} f_n(r_0,\phi_0) &= q(t) f_n(r(t),\phi(t)).
\end{align}
If we take the derivative of both sides with respect to $t$, we have 
\begin{equation*}
P_+e^{tP_+}f_n(r_0,\phi_0) = \frac{d}{dt} \left[  q(t) f_n(r(t),\phi(t)) \right].
\end{equation*}
Expanding both sides of this equation gives
\begin{equation*}
qe^{i\phi}\left(\frac{\partial f_n}{\partial r} + \frac{i}{r} \frac{\partial f_n}{\partial \phi} \right) = f_n \frac{dq}{dt} + q\left( \frac{\partial f_n}{\partial r} \frac{dr}{dt}+\frac{\partial f_n}{\partial \phi}\frac{d\phi}{dt} \right).
\end{equation*}
By equating the derivates, we see that this last equation is true if and only if
\begin{align*}
\frac{dr}{dt} &= e^{i\phi} \\
\frac{d\phi}{dt} &= \frac{ie^{i\phi}}{r} \\
\frac{dq}{dt} &= 0.
\end{align*}
This coupled system of first order differential equations can be solved with a computer algebra package such as Maple.  After solving the system you find that
\begin{align*}
r(t) &= \sqrt{2r_0\phi_0 t + r_0^2 } \\
\phi(t) &= \frac{r_0\phi_0}{\sqrt{2r_0\phi_0 t + r_0^2 }} \\
q(t) &= 1
\end{align*}
Hence, (\ref{eq:second gen func initial}) can be written
\begin{equation*}
e^{tP_+} f_n(r_0,\phi_0) = f_n\left(\sqrt{2r_0\phi_0 t + r_0^2 },\frac{r_0\phi_0}{\sqrt{2r_0\phi_0 t + r_0^2 }}\right).
\end{equation*}
Setting this expression equal to (\ref{bessel gen func first step}) and removing the subscripts from the variables gives us another Bessel generating function:
\begin{equation*}
f_n\left(\sqrt{2r\phi t + r^2 },\frac{r\phi}{\sqrt{2r\phi t + r^2 }}\right) =  \sum_{m=0}^\infty \frac{(-1)^m t^m}{m!} e^{i(n+m)\phi} J_{n+m}(r),
\end{equation*}
or equivalently,
\begin{equation*}
\exp\left( \frac{r\phi}{\sqrt{2r\phi t + r^2 }}\right) J_n\left(\sqrt{2r\phi t + r^2 }\right) = \sum_{m=0}^\infty \frac{(-1)^m t^m}{m!} e^{i(n+m)\phi} J_{n+m}(r)
\end{equation*}
which is (\ref{eq:bessel gen func 2}).

\end{document}